\documentclass[%
 reprint,
 amsmath,amssymb,
 aps,
]{revtex4-2}

\usepackage{graphicx}
\usepackage{dcolumn}
\usepackage{bm}

\usepackage[dvipsnames]{xcolor}

\begin{document}

\preprint{APS/123-QED}

\title{A new generation of reduction methods for networks of neurons with complex dynamic phenotypes}

\author{Inês C. Guerreiro}
\affiliation{%
Group for Neural Theory, Laboratoire de Neurosciences Cognitives Computationnelles INSERM U960, Département d'études cognitives, École Normale Superieure, Paris Sciences \& Lettres University, Paris 75005, France 
}%

\author{Matteo di Volo}
\affiliation{%
Université Claude Bernard Lyon 1, Institut National de la Santé et de la Recherche Médicale,
Stem Cell and Brain Research Institute U1208, Bron, France
}%

\author{Boris Gutkin}
\affiliation{%
Group for Neural Theory, Laboratoire de Neurosciences Cognitives Computationnelles INSERM U960, Département d'études cognitives, École Normale Superieure, Paris Sciences \& Lettres University, Paris 75005, France 
}%
\affiliation{
Center for Cognition and Decision Making, Institute for Cognitive Neuroscience, National Research University Higher School of Economics, Moscow 101000, Russia
}%

\date{\today}

\begin{abstract}

Collective dynamics of spiking networks of neurons has been of central interest to both computation neuroscience and network science. Over the past years a new generation of neural population models based on exact reductions (ER) of spiking networks have been developed. However, most of these efforts have been limited to networks of neurons with simple dynamics (e.g. the quadratic integrate and fire models with periodic firing). Here, we present an extension of ER to conductance-based networks of two-dimensional Izhikevich neuron models.  We employ an adiabatic approximation, which allows us to analytically solve the continuity equation describing the evolution of the state of the neural population and thus to reduce model dimensionality. We validate our results by showing that the reduced mean-field description we derived can qualitatively and quantitatively describe the macroscopic behaviour of populations of two-dimensional QIF neurons with different electrophysiological profiles (regular firing, adapting, resonator and type III excitable). Most notably, we apply this technique to develop an ER for networks of neurons with bursting dynamics 

\end{abstract}

\maketitle


\section{Introduction}

For decades, theoretical neuroscientists have been using mean-field theory to reduce the description of neural circuits composed of many interacting neurons to a low-dimensional system that describes the macroscopic dynamical states of the network.  This mean-field approach is quite powerful as it generates a reduced picture of the neural population that can be used to study how brain functions arise from the collective behaviour of spiking neurons. Moreover, these mean field models are today largely employed as a building block to construct large-scale models of the brain \cite{sanz2013virtual} and studying cognitive process \cite{mejias2022mechanisms}.Different techniques have been employed in recent years based on approximations of the underlying network dynamics \cite{augustin2017low,schwalger2017towards,diVolo,Nicola,di2021optimal}. Recently, an exact reduction method based on the Ott Antonssen ansatsz \cite{ott2008low} has been introduced that allows to derive exact mean-field  equations for heterogeneous and globally coupled networks of quadratic integrate-and-fire (QIF) neurons \cite{Montbrio}. This methodology has been employed to study networks with electrical synapses \cite{montbrio2020exact}, delays \cite{pazo2016quasiperiodic}, working memory \cite{taher2020exact}, to analytically estimate cross-frequency-couplings \cite{dumont2019macroscopic} and recently to study brain activity at the whole brain level \cite{gerster2021patient}. Moreover, recent work shows that it is possible to derive mean field equations also for sparse networks in presence of noise \cite{goldobin2021reduction}. However, so far the exact-reduction approach has been limited to networks of one-dimensional quadratic integrate-and-fire (QIF) neurons that cannot account for complex spiking and bursting dynamics.

Recent advances using two dimensional models with spike frequency adaptation have been developed, but also in this case neurons do not show intrinsic bursting dynamics \cite{ferrara2023population}.

On the other hand, two-dimensional quadratic integrate-and-fire models (e.g., the Izhikevich neuron model \cite{Izhikevich}) reproduce a wide variety of spiking and bursting behaviours.  Yet reduced descriptions of networks that cover the various Izhikevich neuron dynamical profiles have been a largely lacking. A mean-field reduction of such neuron models will enable us to derive the macroscopic dynamics of populations of neurons with a wide variety of spiking properties. 

In this work, we propose a reduction method for  large networks of conductance-based interacting Izhikevich neurons. We start by presenting the two-dimensional  neuron model and then show how a separation of time scales of the variables describing the state of the neurons allows us to explicitly solve the continuity equation of the system, which represents the evolution of the state of the neural population. By doing so, we obtain a system of two coupled variables, the firing rate and the mean voltage, which together describe the evolution of the macroscopic system.  We study the accuracy of the neural mass by comparing its predictions with numerical  simulations of finite-size networks of neurons with different spiking properties (e.g. regular firing, adapting, bursting) and excitability types. Finally, we show that our reduction approach can be employed to derive a neural mass model of interacting intrinsically bursting neurons. All together, our results open the possibility of generating realistic mean-field models from electrophysiological recordings of individual neurons and can be used to relate the biophysical properties of neurons with emerging behaviour at the network scale.

\section{Population model of Izhikevich neurons }

We derive a mean-field model for populations of coupled Izhikevich neurons. Each neuron $i$ from a population $Y$ is described by a fast variable representing the membrane potential, $V\ (mV)$, and a slow variable representing the recovery current, $u\ (pA)$:

\begin{subequations}
\label{eq:all_QIF}  
\begin{align}
& C_m \frac{dV_i^Y}{dt} = a(V_i^Y - V_r)(V_i^Y - V_t) - u_i^Y + I_i \label{eq:V_QIF}\\   
& \frac{du_i^Y}{dt} = \alpha (\beta (V_i^Y - V_r) - u_i^Y)
\end{align}
\end{subequations}

where the onset of an action potential is taken into account by a discontinuous reset mechanism:

\begin{align*}
If\ V_i^Y>V_{peak} \Rightarrow V_i^Y \leftarrow V_{reset},\  u_i^Y \leftarrow u_i^Y + u_{jump}
\end{align*}

The parameters are as follows: $C_m$ stands for the membrane capacitance, $V_r$ is the resting potential, $V_t$ the threshold potential, $a$ is a scaling factor, $\alpha$ the time constant of the recovery variable $u$, $\beta$ modulates the sensitivity of the recovery current to subthreshold oscillations, and $I_i$ is the total current acting on neuron $i$. We consider $I_i =  \eta_i + I_{ext} + I_{syn,i}$. The parameter $\eta_i$ represents a background current. To account for the network heterogeneity, the parameter $\eta_i$ is randomly drawn from a Lorentzian distribution with half-width $\Delta$ centered at $\overline{\eta}$, $g(\eta) = \frac{1}{\pi} \frac{\Delta}{(\eta - \overline{\eta})^2 + \Delta^2}$. $I_{ext}$ is an external current acting on population W (identical to all neurons).  $I_{syn,i}$ is the total synaptic current acting on neuron $i$ given by:

\begin{equation}
I_{syn,i} = \sum_Z s_{YZ} (E_r^Z - V_i^Y)
\end{equation}

where $E_r^Z$ is the reversal potential of the synapse, and $s_{YZ}$ the synaptic conductance. If we assume that all neurons of population $Y$ are connected to all neurons of population $Z$, the synaptic conductance $s_{YZ}$ can be described according to the following equation:

\begin{equation}
\frac{ds_{YZ}}{dt} = -\frac{s_{YZ}}{\tau_s} + \frac{p_{YZ}}{N_Z}\sum_{k=1}^{N_Z} \sum_f \delta(t - t_f^k)
\end{equation}

where $\delta$ is the Dirac mass measure and $t_f^k$ is the firing time of neuron $k$. The parameter $\tau_s$ represents the synaptic time constant, $N_Z$ is the number of neurons of population $Z$, and $p_{YZ}$  is the coupling strength of population $Z$ on population $Y$.

\subsection{\label{sec:ad_app}Adiabatic approximation of the two-dimensional 
Izhikevich neuron model}

We exploit the time scales difference between the dynamics of the membrane potential $V$ and the recovery variable $u$ to reduce the dimensionality of the neural network. If the time scale of the recovery variable is much slower than the other variables, we can invoke an adiabatic approximation by considering that all neurons of population $Y$ receive a common recovery variable $u^Y$. This results in the modified Izhikevich QIF model:

\begin{subequations}
\begin{align}
C_m \frac{dV_i^Y}{dt} = & a(V_i^Y - V_r)(V_i^Y - V_t) - u^Y +  I_i \label{eq:Vi_ad}\\   
\frac{du^Y}{dt} = & \alpha(\beta(\langle V \rangle ^Y - V_r) - u^Y) \nonumber \\
& + u_{jump} \sum_{k=1}^{N_Y} \sum_f \delta(t - t_f^k)  \label{eq:u_ad} 
\end{align}
\end{subequations}

where $\langle V \rangle ^Y$ is the mean membrane potential of the population $Y$, described as follows:

\begin{equation}
\langle V \rangle ^Y = \frac{\sum_{k=1}^{N_Y} V_k^Y}{N_Y}
\label{eq:av_V}
\end{equation}

Note that we have incorporated the resetting mechanism of the variable $u^Y$ in the last term of equation \eqref{eq:u_ad}. The onset of an action potential is now described by:

\begin{equation*}
V_i^Y > V_{peak} \Rightarrow V_i^Y \leftarrow V_{reset}
\end{equation*}

From now on we will consider equation \eqref{eq:Vi_ad} written in terms of the parameters $b=a(-V_r - V_t)$ and $c=aV_rV_t$:

\begin{align}
C_m \frac{dV_i^Y}{dt} = & a(V_i^Y)^2 + bV_i^Y +c - u^Y +   \eta_i \nonumber \\
&+\sum_Z s_{YZ}(E_r^Z - V_i^Y) + I_{ext}
\end{align} 

The main consequence of the adiabatic approximation is the reduction in the number of state variables describing a neuron in the population from $(V_i^Y, u_i^Y)$ to $(V_i^Y)$. This is a crucial step in our method since it enables us to solve the continuity equation of the system analytically, as we demonstrate in the next section.

\section{\label{sec:mean_field}Mean-field reduction}

Having reduced the dimensionality of the Izhikevich neuron model, in the adiabatic approximation we have that the effective current acting on neuron $i$ is $I_i\to I_i+u^Y$. As a result, we can employ the reduction method developed in  \citep{Montbrio} for a one-dimensional QIF neuron model, thus obtaining a macroscopic description of a population of Izhikevich neurons. The reduced model obeys the following differential equations  (see appendix \ref{ap:MF} for more details):
 
 \begin{subequations}
\label{all:mean_field_princ}
\begin{align}
 C_m \frac{dr_Y}{dt} &= (b - \sum_Z s_{YZ})r_Y + 2ar_Yv_Y + \Delta\frac{a}{C_m \pi} \label{eq:r_final}\\
C_m \frac{dv_Y}{dt} &= -\frac{C_m^2 \pi^2}{a}r_Y^2 + av_Y^2 + c  - u^Y + (b - \sum_Z s_{YZ})v_Y\nonumber\\
& + I_{ext} + \sum_Z E_r^Z + \overline{\eta} \label{eq:v_final}\\
 \frac{du^Y}{dt} &= \alpha (\beta(v^Y - V_r) - u^Y) + u_{jump}r_Y
\label{eq:mean-field}
\end{align}
\end{subequations}

with 

\begin{equation}
\frac{ds_{YZ}}{dt} = - \frac{s_{YZ}}{\tau_s} + p_{YZ}r_Z.
\label{eq:u_mean-field}
\end{equation}

Here $r_Y$ is the population firing rate, $v_Y$ the mean membrane potential across neurons and $u^Y$ the mean slow recovery variable (e.g. adaptation)

\subsection{Numerical simulations for multiple dynamic phenotypes}
\label{subsec:sim}
The Izhikevich two-variable QIF model can, with the adequate choice of parameters, reproduce many of the key features of firing patterns observed in neurons, such as tonic spiking, subthreshold oscillations, and bursting \citep{Izhikevich}. We here compare the neural mass model of Eq.s \ref{all:mean_field_princ} with direct simulations of Izhikevich neurons in different parameter regimes characterized by different intrinsic firing dynamics of neurons.

\begin{figure*}[htbp]
\begin{center}
  \centerline{ \includegraphics[width=\textwidth]{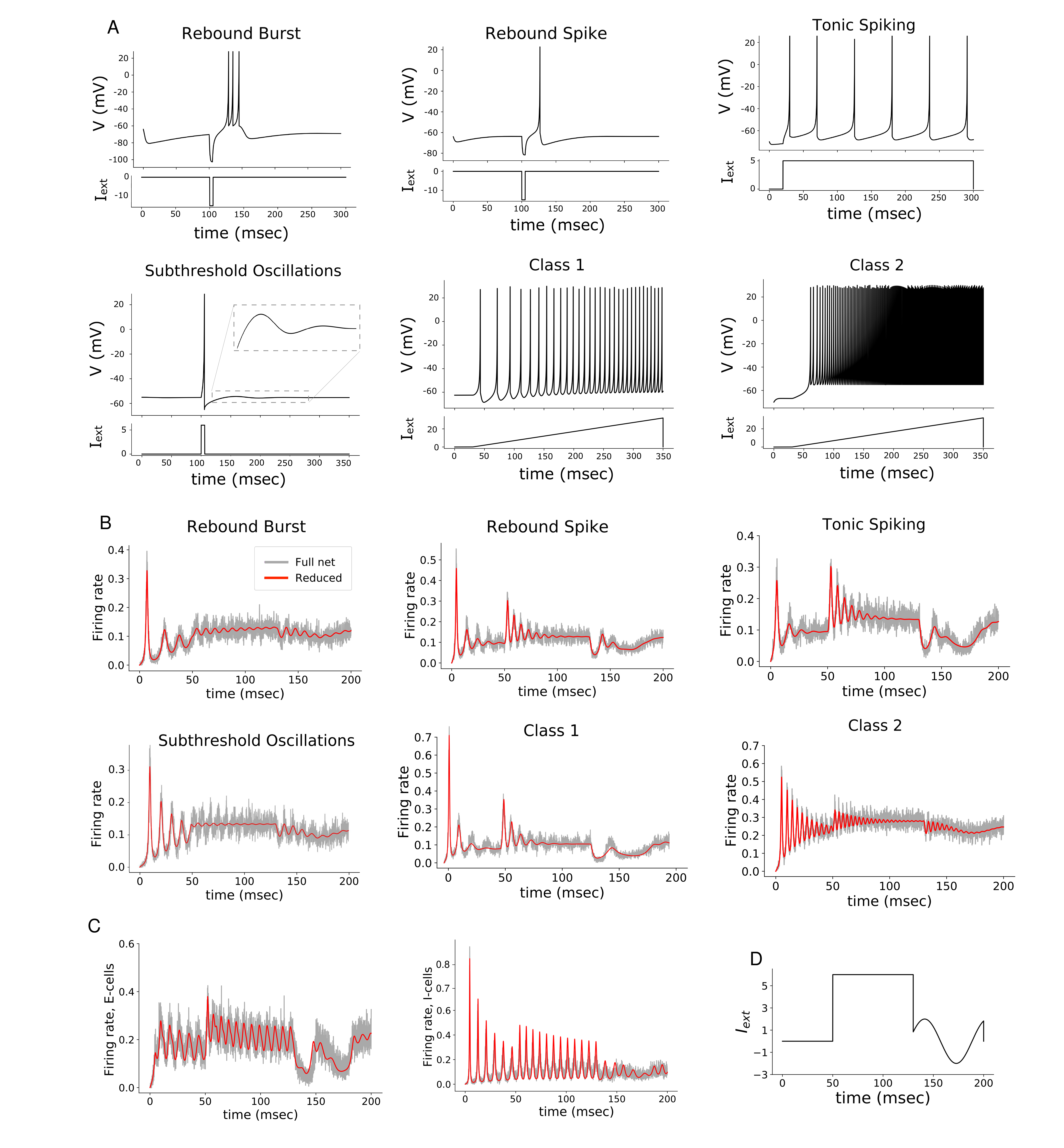} }
\caption[Comparison between the full network and exact reduced system for networks of neurons with distinct dynamics. ]{\textbf{Comparison between the full network and reduced system for networks of neurons with distinct dynamics.} \textbf{(A)} Membrane potential of spiking neurons with different spiking features. Results were obtained using the Izhikevich two-dimensional QIF neuron model \cite{Izhikevich} with the adequate choice of parameters (see Table \ref{tab:param_QIF}). \textbf{(B)} Firing rate of populations of uncoupled neurons with different dynamics obtained from simulations of the full and reduced system, and respective external current.\textbf{(C)} Firing rate of a population of recurrently connected excitatory tonic spiking cells (E) and inhibitory neurons with subthreshold oscilations (I). \textbf{(D)} External current acting on all neuronal populations. Parameters: $\Delta=1$, $\overline{\eta}=15$, $N=3000$, $p_{EE}=1$, $p_{EI}=p_{IE}=1$, $p_{II}=2$, $\tau_s=1$.  }
\label{fig:comp_FullRed}
\end{center}
\end{figure*}  
  
Figure \ref{fig:comp_FullRed} illustrates a comparison of the dynamics of the full network of Izhikevich QIF neurons with its corresponding reduced system. Regarding the full system, each population is made up of $N=3000$ neurons. The neurons are described by the two-dimensional QIF model \ref{all:QIF}, with the respective parameters specified in Table \ref{tab:param_QIF}. The firing rate is calculated according to:$ r(t) = \frac{1}{N}\sum_{k=1}^{N}\sum_f \delta(t - t_f^k)$. For the reduced system, the firing rate is calculated according to equation \ref{eq:r_final}. The reduced description closely follows the firing activity of the full network for all populations.
  
\begin{table}
\begin{ruledtabular}
\begin{tabular}{cccccccc}
                                                & \multicolumn{2}{c}{\textbf{Rebound}} & \multicolumn{2}{c}{\textbf{Tonic}} & \multicolumn{2}{c}{\textbf{Class}} & \multicolumn{1}{l}{}           \\  
                                                & \textbf{Burst}         & \textbf{Spike}        & \textbf{Burst}     & \textbf{Spike}     & \textbf{1}            & \textbf{2}           & {\textbf{Sub. Osc.}} \\ \hline
\multicolumn{1}{l}{a}          & 0.04          & 0.04         & 0.04         & 0.04        & 0.04         & 0.04        & \multicolumn{1}{c}{0.0454}    \\
\multicolumn{1}{l}{b}             & 5.3           & 4.99         & 4.88         & 4.93        & 4.96         & 4.98        & \multicolumn{1}{c}{5.02}      \\
\multicolumn{1}{l}{c }        & 174           & 154          & 148.2        & 152         & 154          & 155         & \multicolumn{1}{c}{137.76}    \\
\multicolumn{1}{l}{$C_m$ }      & 1             & 1            & 1            & 1           & 1            & 1           & \multicolumn{1}{c}{2}         \\
\multicolumn{1}{l}{$V_r$}                & -65           & -56          & -65          & -60         & -65          & -65         & \multicolumn{1}{c}{-60}       \\
\multicolumn{1}{l}{$\alpha$}    & 0.01          & 0.03         & 0.02         & 0.02        & 0.02         & 0.2         & \multicolumn{1}{c}{0.05}      \\
\multicolumn{1}{l}{$\beta$}       & 0.9           & 0.25         & 0.32         & 0.2         & 0.1          & 0.26        & \multicolumn{1}{c}{1.1}       \\
\multicolumn{1}{l}{$u_{jump}$} & 0             & 4            & 0            & 2           & 4            & 0           & \multicolumn{1}{c}{0}         \\
\multicolumn{1}{l}{$V_{peak}$}           & 30            & 30           & 30           & 30          & 30           & 30          & \multicolumn{1}{c}{30}        \\
\multicolumn{1}{l}{$V_{reset}$ }          & -60           & -60          & -55          & -60         & -60          & -55         & \multicolumn{1}{c}{-55}       \\
\end{tabular}
\end{ruledtabular}
\caption{Parameter values of two-dimensional QIF neuron model for neurons displaying different firing properties. Parameters adapted from \cite{Izhikevich}. a[$mS/cm^2 mV$], b[$mS/cm^2$],c[$mS/(cm^2 mV)$], $C_m$[$\mu F/cm^2$], $V_r$[mV], $\alpha$[$msec^{-1}$], $\beta$[$mS/cm^2$], $u_{jump}$[$\mu A/cm^2$], $V_{peak}$[mV], $V_{reset}$[mV] }
\label{tab:param_QIF}
\end{table}

\subsection{Limitations of reduction formalism}
\begin{figure}[htbp]
\begin{center}
\label{fig:error_Class1}
  \centerline{ \includegraphics[width=\columnwidth]{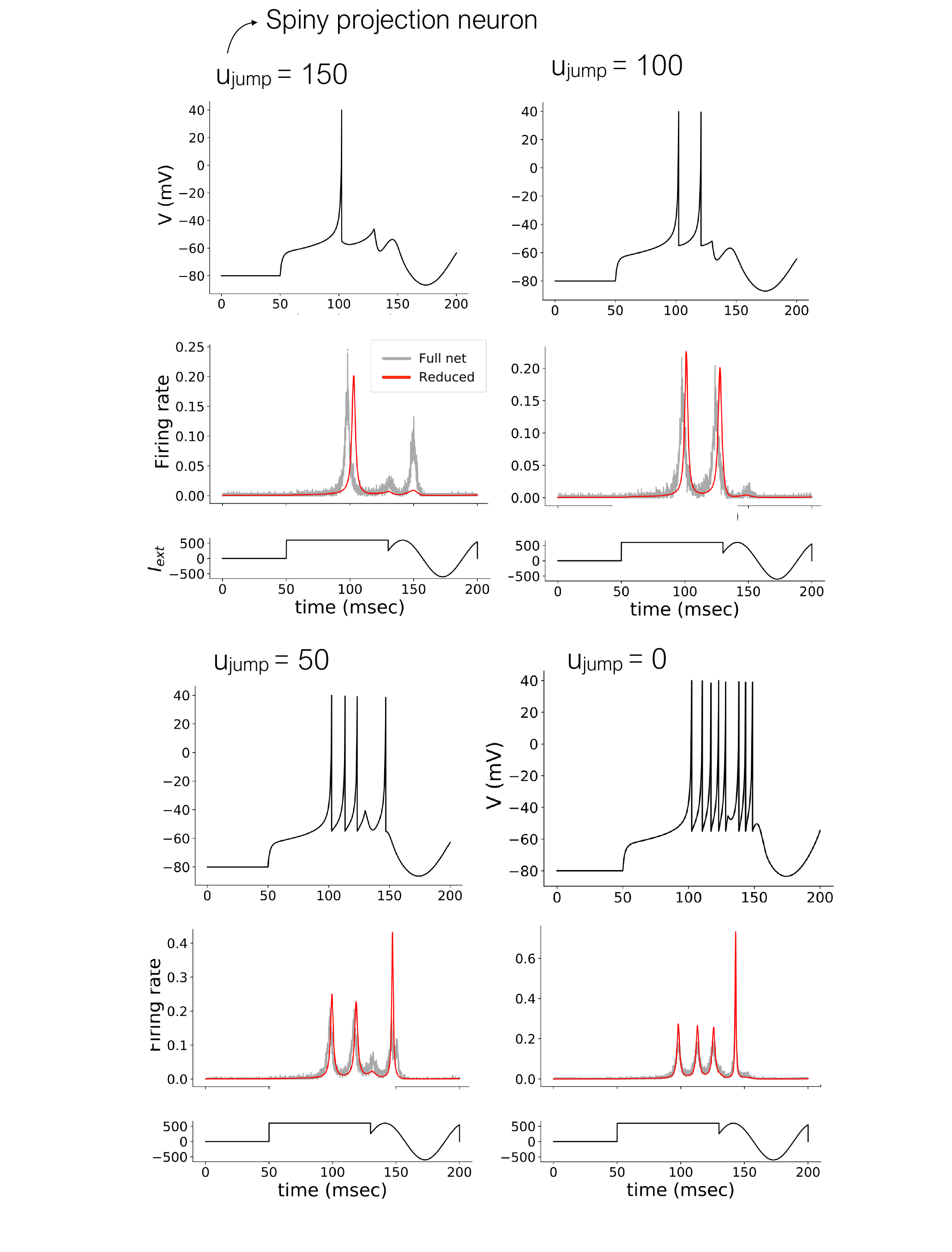} }
\caption[Comparison between full and reduced system for a population of spiny projection neurons.  ]{ \textbf{Comparison between full and reduced system for a population of spiny projection neurons.} Spiny projection neurons of the neostriatum and basal ganglia can be described by the two-dimensional QIF neuron model with $a=1\ nS/mV$, $b=105\ nS $, $c=2000\ nS mV $, $C_m=50\ pF$, $V_r=-80\ mV$, $\alpha=0.01\ msec^{-1}$, $\beta=-20\ nS$, $V_{peak}=40\ mV$, $V_{reset}=-55\ mV$ and $u_{jump}=150\ pA$ \citep{Iz_book}. Decreasing the value of $u_{jump}$ improves representation of the population activity.}
\label{fig:spiny}
\end{center}
\end{figure}

One crucial assumption of the mean-field reduction formalism here presented is that the recovery variable $u$ follow sufficiently slow dynamics. However, every time a neuron reaches $V_{peak}$, the membrane potential $V$ is reset to $V_{reset}$ and the recovery variable $u$ is instantaneously increased by $u_{jump}$, adding a discontinuity to the system. Given that the adiabatic approximation made in section \ref{sec:ad_app} relies on the assumption that the variable $u$ is the slowest variable in the system and that therefore we can consider that all the neurons in the population receive a variable $u$ with approximately the same value, adding an instantaneous jump brings the approximation into question. The larger the jump, the more evident this point becomes. This the spike-dependent jump of the recivery variable $u$ should be sufficiently small in order for the approximation to work well. In the examples considered in Figure \ref{fig:comp_FullRed}, $u_{jump}$ is sufficiently small for the reduction to work with minimal errors. This means the reduced system here derived can be used to study the activity of populations with any of the spiking dynamics portrayed in Figure \ref{fig:comp_FullRed}. Still, it may not be adequate to describe the activity of certain neural populations, such as rat spiny projection neurons of the neostriatum and basal ganglia \citep{Iz_book}, for which the value of $u_{jump}$ used to describe their dynamics is large enough to induce imprecisions (Figure \ref{fig:spiny}).

In Figure \ref{fig:spiny} we compare the full and reduced system for a population of uncoupled spiny projection neuron models ($u_{jump}=150$ pA) \citep{Iz_book}. 
We then systematically decrease the value of $u_{jump}$ we see how that changes the accuracy between dynamics of the full and reduced system. All the neurons receive an external current as described in Figure \ref{fig:comp_FullRed} (D). We see that there is not a perfect agreement between the full and reduced system for a population of spiny projection neurons (left panel). Decreasing the value of $u_{jump}$ notably improves the agreement between the full and reduced system significantly, confirming that the high $u_{jump}$ is at the origin of the mismatch observed. For $u_{jump}=150$, one way to improve the representation of the population activity would be to decrease $\Delta$. By decreasing the variance $\Delta$ of the intrinsic variable $\eta$, we decrease the heterogeneity of the network. As a result, we can consider again that all the neurons in a population $W$ are receiving the same variable $u$ at any given time.

\subsubsection{The particular case of bursting neurons}
\begin{figure*}[htbp]
\begin{center}
  \centerline{ \includegraphics[width=0.7\textwidth]{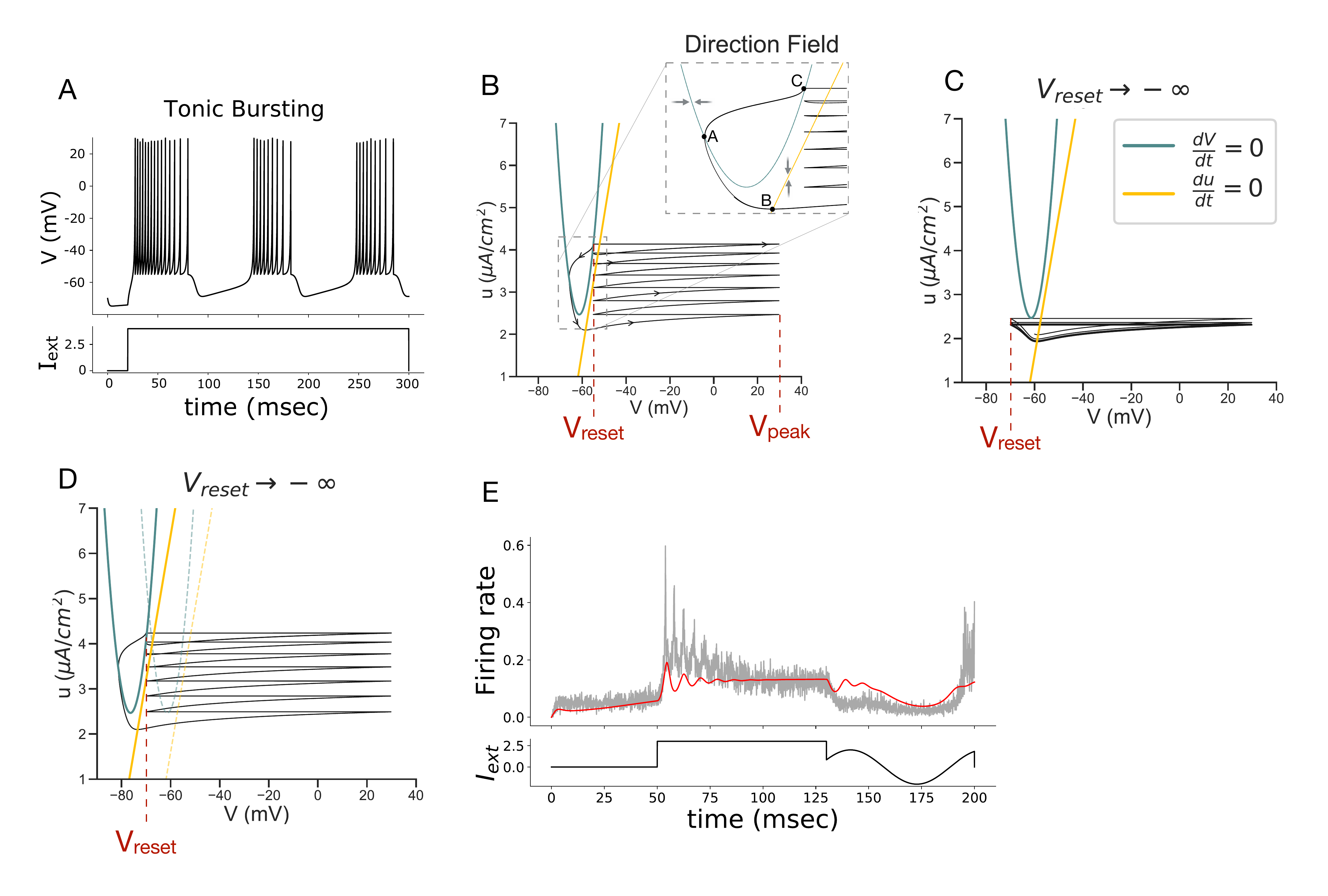} }
\caption[ ]{ \textbf{Comparison of reduced and full system for a class of bursting neurons.} \textbf{(A)} Voltage trace of a bursting neuron using the Izhikevich QIF neurons model. \textbf{(B)} Nullclines, $\frac{dV}{dt}=0$ (green line) and $\frac{du}{dt}=0$ (yellow line), for a system of a bursting neuron and respective trajectory (black line) on the phase plane. \textbf{(C)} Nullclines and trajectory of the system when $V_{reset}$ decreases from -55 to -70 mV on the phase plane. The trajectory of the system no longer shows a bursting behavior. \textbf{(D)} Nullclines and trajectory of the system on the phase plane when $b=6$, $c=232$, $V_r=-80$ and $V_{reset}=-70$. As a result of the changes in $b$, $c$ and $V_r$ the nullclines moved to the left of the phase plane and we recover the trajectory of bursting neurons. \textbf{(E)} Comparison between full and reduced system for a population of bursing neurons (with $b=6$, $c=232$, $V_r=-80$ and $V_{reset}=-70$). The reduced system captures some but not all of the structure of the full bursting system. }
\label{fig:bursting}
\end{center}
\end{figure*}

A critical point of the derivation of our reduced mean-field model is the assumption that the firing rate of a population is defined as the flux $J(V,t)(= \frac{dV}{dt}\rho(V,t))$ at infinity. In other words, we consider $V_{peak} \rightarrow \infty$. Similarly, for the reduction we assume that $V_{reset} \rightarrow - \infty$. While moving $V_{peak}$ towards infinity does not change the intrinsic spiking properties of the neurons that constitute the population, moving $V_{reset}$ in the direction of $-\infty$ changes the microscopic behavior of bursting neurons.
We can clearly see how the bursts are built by looking  at the phase portrait of an intrinsically bursting Izhikevich neuron in Figure \ref{fig:bursting} (B). Starting at point A, we are on the V-nullcline, where by definition $\frac{dV}{dt} = 0$, and the dynamics is going to be governed by the u-component. Since we are on the left of the u-nullcline, the trajectory follows a downward flow. As $u$ slowly decreases, we reach point B below the V-nullcline, and the fast dynamics in the $V$ direction pushes the system towards $V_{peak}$, at which point the system is reset to $V_{reset}$. This last process repeats while $u$ slowly increases until it reaches point C, where a voltage reset takes the system to a point above the V-nullcline. In this region, the flux is directed towards the left, which brings the system back to point A.

By decreasing the value of $V_{reset}$, we lose the bursting dynamics, and the neuron model now shows tonic spiking instead (see Figure \ref{fig:bursting} (C)). One way to preserve the bursting dynamics of the microscopic system would be to move the V and u-nullclines by the same amount as the $V_{reset}$ (Figure \ref{fig:bursting} (D)). We do so by decreasing the values of $V_r$ and $V_t$ (remember that $b=-a(V_r + V_t)$ and $c=aV_rV_t$). From Figure \ref{fig:bursting} (E), we see that by adopting this change the full and reduced system activity have approximately the same shape, but that they do not perfectly agree. It is important to note, however, that this method presents important faults: it implies that at $V_{reset} \rightarrow - \infty$, the resting and threshold potential, $V_r$ and $V_t$, should also move to $-\infty$. This is not only a problem from the biological point of view, but it can also invalidate some mathematical results adopted during the derivation of the mean-field reduction; namely, when solving explicitly the integrals that define the firing rate and mean voltage of the population (equations \eqref{eq:r2} and \eqref{eq:v2}).

An alternative solution is to consider the two-dimensional theta neuron model with a slow recovery variable, which with the appropriate choice of parameters can produce bursting \citep{Erm_Kop}, and apply the mean-field reduction. In the theta neuron model, the system evolves along a circle and $V \in [-\infty, +\infty]$ maps to $\theta \in [0, 2\pi]$. We note that we can construct a two-dimensional theta-neuron that is mathematically equivalent to the Izhikevich model.

An example of a bursting theta neuron model is the following:

\begin{subequations}
\begin{align}
&\frac{d\theta^Y_i}{dt} = 2( 1 - cos(\theta^Y_i/2) + (I_{ext} + \eta_i - u^Y_i)(1 + cos(\theta^Y_i/2))   )\\
& \frac{du^Y_i}{dt} = \alpha ( \beta(  1 + \frac{tan(\theta^Y_i/4)}{1 + 2( 1 + tan(\theta^Y_i/4)^2  )}  ) - u^Y_i )
\label{eq:theta_neuron}
\end{align}
\end{subequations}

with $I_{ext}=1.45$, $\alpha=0.1$ and $\beta=1.39$.

\begin{figure}[htbp]
\begin{center}
  \centerline{ \includegraphics[width=0.6\columnwidth]{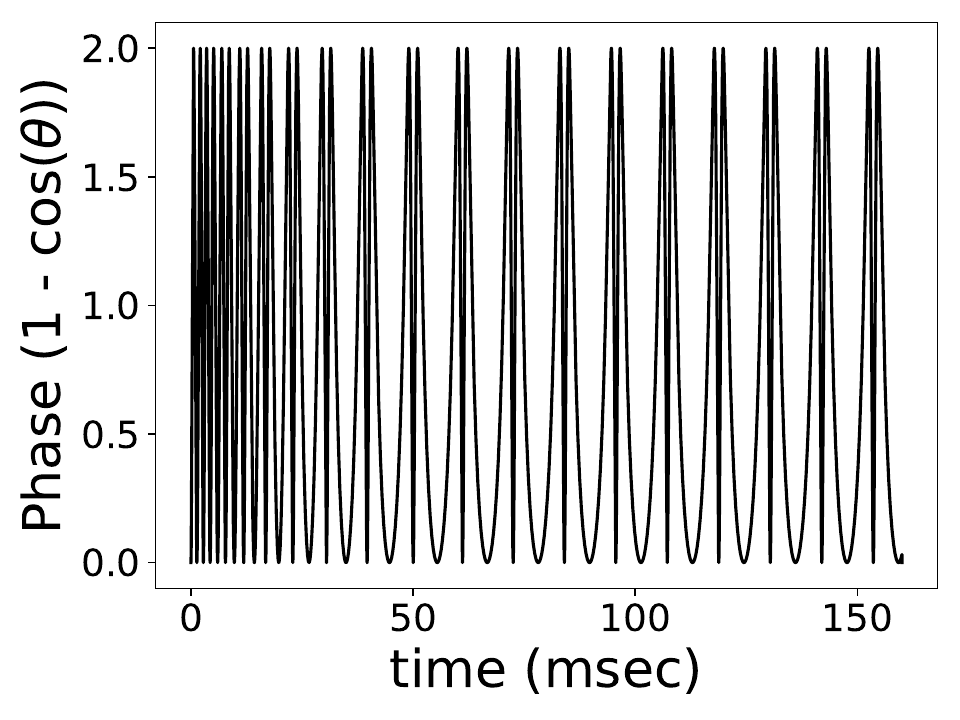} }
\caption[Phase of a two-variable theta neuron model]{\textbf{Phase of a two-variable theta neuron model} Figure reproduced using equations \eqref{eq:theta_neuron} with $I_{ext}=1.45$, $\alpha=0.1$ and $\beta=1.39$ }
\label{fig:comp_theta_burst}
\end{center}
\end{figure}

If we consider a population of theta neurons described by equation \eqref{eq:theta_neuron}, where $\eta_i$ for each neuron $i$ of a population $Y$ is randomly drawn from a Lorentzian distribution with half-width $\Delta$ centered at $\bar{\eta}$,  $g(\eta) = \frac{1}{\pi} \frac{\Delta}{(\eta - \overline{\eta})^2 + \Delta^2}$, we can employ the reduction method to obtain mean-field equations

\begin{subequations}
\label{eq:burst_MF}
\begin{align}
\frac{d|Z|}{dt} & = \frac{1}{2}(   sin(\theta^Y/2)(\bar{\eta} + I_{ext} - u^Y - 1) \nonumber \\
& - \Delta cos(\theta^Y/2) (1 + |Z|^2) - 2|Z|\Delta \nonumber \\
& + |Z|^2 sin(\theta^Y/2)(1 - \bar{\eta} - I_{ext} + u^Y)   ) \label{eq:whatever} \\
\frac{d\theta^Y}{dt} & = \frac{cos(\theta^Y/2)}{2}(\bar{\eta} + I_{ext} - u^Y - 1) \nonumber \\
& + \frac{\Delta}{2}sin(\theta^Y/2)(1 - |Z|^2) + 2(\bar{\eta} + I_{ext} - u^Y) \nonumber \\
& + 2 + |Z|cos(\theta^Y/2)(\bar{\eta} + I_{ext} - u^Y - 1) \label{eq:whatever2}\\
\frac{du^Y}{dt}  & = \alpha( \beta(1 - \frac{B( 3-2\sqrt{6}A + 2A^2 + 2B^2 )}{8B^2 (2 + B^2) + (3 - 2A^2)^2}) -u^Y) 
\label{eq:wathever3}
\end{align}
\end{subequations}

with $A = \frac{1 - |Z|^2}{1 + |Z|^2 + 2|Z| cos(\theta^Y/2)}$ and $B = \frac{ -2|Z|sin(\theta^Y/2)}{ 1 + |Z|^2 + 2|Z| cos(\theta^Y/2) }$. Here, the evolution of the population of theta neurons is described in terms of the macroscopic variable $Z = |Z|exp(-i\theta^Y)$, where $\theta^Y$ is the mean phase and $|Z|$ the coherence across neurons (see Appendix \ref{sec:burst_MF} for the detailed derivation of equations \eqref{eq:burst_MF})

\begin{figure*}[htbp]
\begin{center}
  \centerline{ \includegraphics[width=0.6\textwidth]{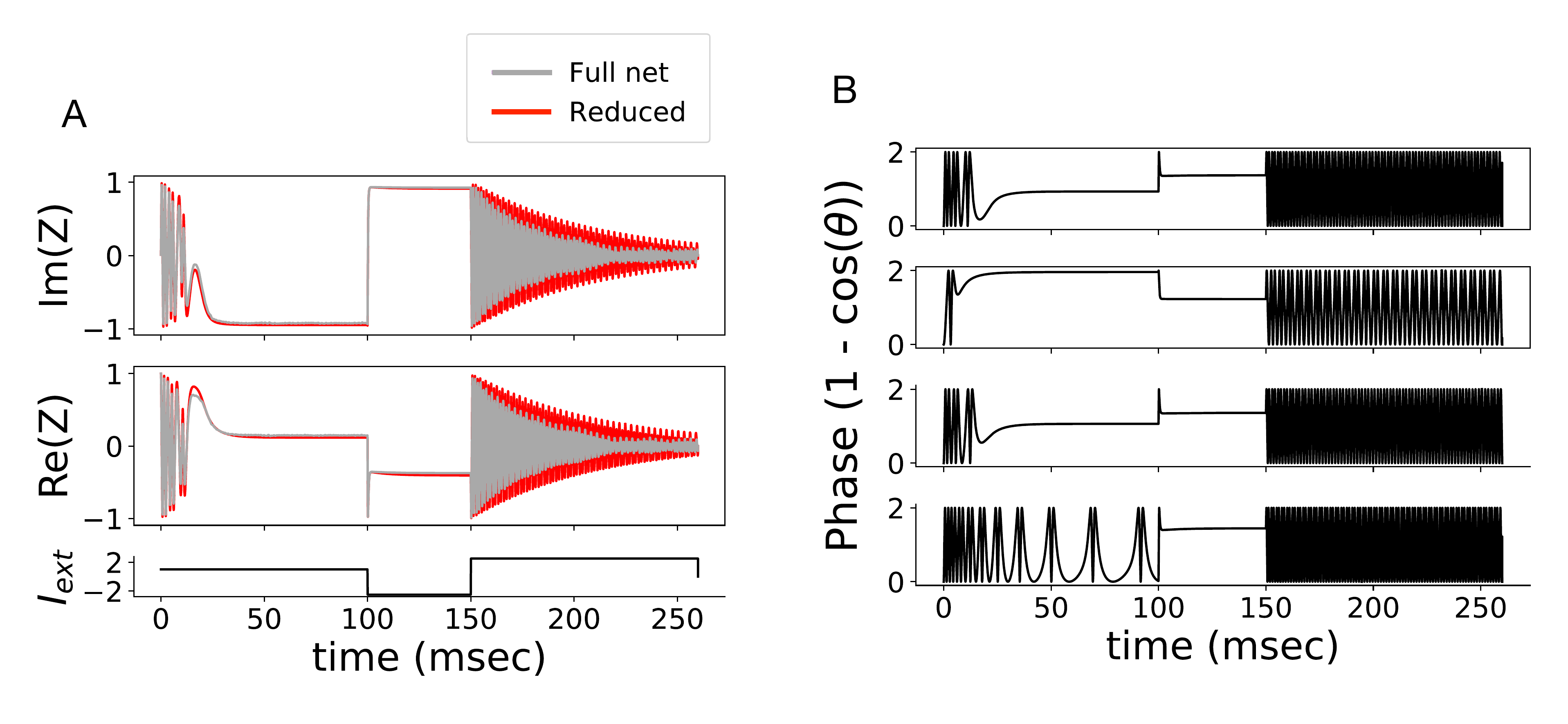} }
\caption[Comparison between the full network and reduced system for a population of theta neurons. ]{\textbf{Comparison between the full network and reduced system for a population of theta neurons} \textbf{(A)} Imaginary and real parts of the macroscopic variable $Z=|Z|exp(-i\theta^Y)$ reproduced by averaging over the entire population using equations \eqref{eq:theta_neuron} (full network), and by using equations \eqref{eq:burst_MF} (reduced). \textbf{(B)} Phase of four randomly selected theta neurons. Some of the neurons in the populations present a bursting-like behavior for a given current $I_{ext}$. Parameters: $\Delta=0.02$, $\overline{\eta}=0.1$, $N=600$ }
\label{fig:comp_MF_burst}
\end{center}
\end{figure*}

\section{Discussion}

In this work, we presented a reduction formalism that allows us to predict the collective large network dynamics of conductance-based interacting spiking neurons with a variety of spiking properties. This was done in two steps. Frist, starting with a population of two-dimensional Izhikevich neurons \cite{Izhikevich}, we employed an adiabatic approximation of the slow recovery variable, which enabled us to reduce the dimension of variables that describes the state of the network. Second, We applied the  Lorentzian ansatz to solve the continuity equation and reduce our full network to a low-dimensional macroscopic system. Notably, we were able to derive population descriptions for neurons that show the following excitability phenotypes: rebound burst and spike, tonic spiking, subthreshold oscillations, and class 1 and 2 of excitable neurons.

Sufficient requirements for our approach to be valid are that the recovery variable $u$ is the slowest in the system and that $u_{jump}$ is relatively small. This means that even though it is possible to describe any class of spiking dynamics, the reduced model might be unable to describe the activity of specific neural populations, such as spiny projection neurons of the neostriatum and basal ganglia whose models require a rather high value of $u_{jump}$ \citep{Iz_book}. 

It is important to note that even though the original Izhikevich neuron model, when put in the appropriate parameter regime, can clearly model tonic bursting neurons, if we were to apply the reduction in a standard manner, the mean-field system would appear to be inadequate to describe the population's behavior - loosing the bursting.  Since, in the Izhikevich two-dimensional QIF model, the bursting mechanism depends on the position the system acquires in the phase-plane ($V$,$u$) upon reset (the reset needs to be above the $V-nullcline$), moving $V_{reset}$ to $\-infty$ alters the behavior of the microscopic system - the bursting is lost. Therefore, despite having a good agreement between the full and reduced system, the population at study is no longer a population of tonic bursting neurons but of tonic spiking neurons. A solution found was to move the u and V-nullclines with $V_{reset}$, so that an action potential will reset the system to the same position in the phase-plane relative to the nullclines and preserve the bursting mechanisms of the original model. We do so by decreasing the resting and threshold potential, $V_r$ and $V_t$ by the same amount as $V_{reset}$. While the full and reduced system of the resultant tonic bursting neurons do not perfectly agree, but the reduced network now accurately reproduces the oscillatory behavior of the population. In other words, when the system receives a strong enough external input $I_{ext}$, both the full and reduced system show damped oscillations yet there is a frequency mismatch. Despite this, we suggest that the mean-field description may still be useful to study certain features of a population of bursting neurons and qualitative behavior. However, it is important to note that the approach taken for the case of the bursting neurons presents some fundamentals problems. Namely, it implies that both the reset, resting and threshold potential are set to $-\infty$. An alternative solution, that we pursued, is to consider the two-dimensional theta neuron model with a slow recovery variable, which with the appropriate choice of parameters can produce bursting \citep{Erm_Kop}, and apply the steps as for the derivation of a two-dimensional QIF. In the theta neuron model, the system evolves along a circle and $V \in [-\infty, +\infty]$ maps to $\theta \in [0, 2\pi]$. In this framework, theta neurons have a periodicity of $2\pi$: whenever the dynamical variable $\theta$ reaches the value $\theta = \pi$, the model is said to produce a spike and to reset to $\theta = 2\pi$. This means that it is not necessary to change the boundary conditions to get a bursting neuron. With the appropriate conformal map, we can make use of the reduction method previously used to obtain a macroscopic description of a population of bursting theta neurons straightforwardly. We have done so by applying the conformal map $W = \frac{1 - Z^*}{1+Z^*}$ to a mean-field description of the modified two-variable Izhikevich model. The reason why we have done this to a modified version of the Izhikevich model was because we needed to ensure that mapping $V \rightarrow \theta$ would result in a bursting dynamics of the variable $\theta$, which was not the case in the original 2-dimensional QIF model. Using this approach, we get a good match between the reduced and full network of theta neurons (see Figure \ref{fig:comp_theta_burst} A), and guarantee that dynamics of the individual neurons that constitute the population at study is preserved.

A similar adiabatic approach appears in \cite{diVolo} and \cite{Nicola}. Di Volo and colleagues propose a mean-field model of spiking neurons with recovery variable by calculating the transfer function (i.e. neurons' stationary firing rate in response to external spike trains) in a semi-analytical way. This approach, however, assumes that neuron dynamics has a stationary firing rate in response to external spike trains, and it does not allow to study populations of neurons whose transfer function is not completely defined with only one variable (i.e. neurons'  stationary firing rate) \cite{diVolo}. Similar to our approach, Nicola and Campbell \cite{Nicola} use moment closure and a steady-state approximation of the recovery variable to write an expression for the population firing rate, defined as the integral of the population density function. However, they cannot apply the Lorentzian ansatz to solve the integral because they don not consider the heterogeneity of the population. Therefore, for some types of networks it won't be possible to be evaluated explicitly the firing rate integral \citep{Nicola}.   An adiabatic approach has been also employed in \cite{gast2020mean} for QIF neurons with spike frequency adaptation, while in \cite{Chen_Capmbell} moment closure is used to get a neural mass model. Nevertheless, in these reduced models the neurons are not intrinsically bursting as in the case of the Izhikevich model we considered here. Other adiabatic approaches in this field are appearing, e.g. to consider slow variables modeling conductance-based ion exchange \cite{bandyopadhyay2021mean}.

In summary,the mean-field formalism we present provides a paradigm to bridge the scales between population dynamics and the microscopic complexity of the physiology of the individual cells, opening the perspective of generating biologically realistic mean-field models from electrophysiological recordings for a variety of neural populations.

\begin{acknowledgments}
MdV, IG and BSG received financial funding from the  ANR Project ERMUNDY (Grant No. ANR-18-CE37-0014).

IG: designed research, carried out research, wrote the paper; MdV: designed research, carried out research, wrote the paper; BSG: designed research, supervised research, wrote the paper
\end{acknowledgments}

\appendix

\section{Mean field reduction}
\label{ap:MF}

In the mean-field limit, a population of neurons can be well represented by the probability density function, $\rho$. This function represents the proportion of neurons that are in a particular state at time $t$. In our case, the state of a neuron is fully described by its membrane potential. 

We note that even though we started with 2-dimensional models for each neuron, the adiabatic approximation allowed us to express the the recovery as a global variable, that depends only on the population voltage and a sum of the spikes arriving at each neuron from the rest of the population, or the firing rate. We will see below that the reduction allows us to come up with the dynamics of the population voltage and the firing rate, which we can simply plug into the equation for $u^W$. Hence we will come up with a 4-dimensional network description.

We denote $\rho(V^W|\eta, t)$ as the probability of finding a neuron from population $W$ with voltage $V$ at time $t$, knowing that its intrinsic parameter is $\eta$. Defining the flux $J(V|\eta,t)(=\frac{dV}{dt}\rho(V|\eta,t))$ as the net fraction of trajectories per time unit that crosses the value $V$, we can write the continuity equation

\begin{equation}
\frac{\partial}{\partial t} \rho(V|\eta,t) = -\frac{\partial}{\partial V} J(V|\eta,t) 
\label{eq:continuity}
\end{equation}

which expresses the conservation of the number of neurons. Note that in integrate-and-fire models, the number of trajectories is not conserved at $V=V_{reset}$ and $V=V_{peak}$. By taking $V_{reset}$ and $V_{peak}$ to infinity, we ensure that the boundary conditions are the same and that the number of trajectories is conserved \footnote{By considering $V_{peak}$ = - $V_{reset}$ = $\infty$, the resetting rule still captures the spike reset as well as the refractoriness of the neurons.}. 
According to the Lorentzian ansatz (LA) \citep{Montbrio}, solutions of the continuity equation \eqref{eq:continuity} for a population of QIF neurons converge to a Lorentzian-shaped function with half-width $x(\eta,t)$ and center at $y(\eta,t)$ of the form:

\begin{equation}
\rho(V^W|\eta,t) = \frac{1}{\pi} \frac{x(\eta,t)}{ [V - y(\eta,t)]^2 + x(\eta,t)^2}
\label{rho}
\end{equation}

We discuss the validity of the LA here applied in appendix \ref{ap:LA}. Here, $x(\eta,t)$ and $y(\eta,t)$ are statistical variables that represent the low dimensional behavior of the probability density function $\rho$. Adopting the LA, we obtain the low dimensional system:

\begin{align}
C_m \frac{\partial x(\eta,t)}{\partial t} &= (b - \sum_Zs_{WZ})x + 2axy\\
C_m \frac{\partial y(\eta,t)}{\partial t} &= -ax^2 + ay^2 + (b - \sum_Zs_{WZ})y + c \nonumber \\
& - u^W  + I_{ext} + \sum_Zs_{WZ}E_r^Z + \eta
\end{align}

that can be written in the complex form as:

\begin{align}
C_m \frac{\partial w(\eta,t)}{\partial t} &= i(-aw^2 + c - u + I_{ext} + \sum_Zs_{WZ}E_r^Z + \eta)\nonumber \\ 
& +  (b - \sum_Zs_{WZ})y) w
\end{align}

with $w(\eta,t) = ix(\eta,t) + y(\eta,t)$

\subsection{The macroscopic variables: firing rate and mean voltage} 
The firing rate is obtained by summing the flux for all $\eta$ at $V=V_{peak}$. Taking $V_{peak} \rightarrow \infty$ the firing rate of a population $W$ is defined as follows

\begin{equation}
r_W(t) = lim_{V \rightarrow \infty} \int  J(V^W|\eta,t)g(\eta)d\eta
\label{eq:r}
\end{equation}

The mean voltage of the population is obtained by integrating the probability density function $\rho$ for all $V$ and $\eta$ values:

\begin{equation}
v_W(t) = \int \int V^W \rho(V^W|\eta,t)g(\eta) dV^W d\eta
\label{eq:v}
\end{equation}

Adopting the solution for the continuity equation \eqref{rho} and inserting it into equations \ref{eq:r} and \ref{eq:v} we have that the phenomenological variables $x$ and $y$ relate with the firing rate, $r$, and mean voltage, $v$, as follows:

\begin{widetext}
\begin{align}
&r_W(t) = \frac{a}{C_m \pi} \int x(\eta,t) (g(\eta)) d\eta \label{eq:r2}\\
&v_W(t) = \int \int \frac{x(\eta,t)}{\pi} \frac{V^W(t)}{(V^W(t) - y(\eta,t))^2 + x(\eta,t)^2} g(\eta)dV^Wd\eta 
\label{eq:v2}
\end{align}
\end{widetext}
 
To avoid indeterminacy of the improper integral, we resort to the Cauchy principal value to evaluate the integral \ref{eq:v2} ($p.v. \int_{-\infty}^{+\infty} h(x)dx = lim_{R \rightarrow \infty} \int_{-R}^{R} h(x)dx$). In the case of a Lorentzian distribution, the principal value is given by $p.v. \int_{-\infty}^{+\infty} \frac{\sigma}{\pi}\frac{x}{(x - x_0)^2 + \sigma^2} dx = x_0$. We then have that the mean voltage is given by:

\begin{align}
v_W(t) &= \int g(\eta) p.v. \int \frac{x}{\pi}\frac{V^W}{(V^W-y)^2 + x^2}    dV^Wd\eta \nonumber \\
       & = \int g(\eta) y d\eta \label{eq:v3}     
\end{align}

As previously mentioned, in the mean-field limit, the probability distribution function $g(\eta)$ is given by

\begin{align*}
g(\eta) = \frac{1}{\pi} \frac{\Delta}{(\eta- \overline{\eta})^2 + \Delta^2} =  \frac{1}{\pi} \frac{\Delta}{( \eta - (\overline{\eta} + i\Delta) )( \eta - (\overline{\eta} - i\Delta))}
\end{align*}

The distribution $g(\eta)$ has poles at $\eta - i\Delta$ and $\eta + i\Delta$, and can be written as

\begin{align*}
g(\eta) = \frac{1}{2\pi i} (\frac{1}{\eta - (\overline{\eta} + i\Delta)} - \frac{1}{\eta - (\overline{\eta} - i\Delta)})
\end{align*}

The integrals in equations \ref{eq:r2} and \ref{eq:v3} are evaluated by closing the integral contour in the complex $\eta$ plane and using the residue theorem. We then have that the firing rate and mean potential relate to the Lorentzian coefficients $x$ and $y$ according to the following expression:

\begin{align}
& r_W(t) = \frac{a}{C_m \pi} x(\overline{\eta} \pm i\Delta, t) \label{eq:r_x} \\
& v_W(t) = y(\overline{\eta} \pm i\Delta, t) \label{eq:v_y}
\end{align}

Given equations \ref{eq:r_x} and \ref{eq:v_y} and noting that 

\begin{align}
C_m \frac{dx(\overline{\eta} \pm i\Delta,t)}{dt} &= (b - \sum_Zs_{WZ})x + 2axy - (\pm \Delta)\\
C_m \frac{dy(\overline{\eta} \pm i\Delta,t)}{dt} &= -ax^2 + ay^2 + c - u  \nonumber \\
 & \qquad + (b - \sum_Zs_{WZ})y + I_{ext} + \overline{\eta}
\end{align}

we have that the continuity equation reduces to the low-dimensional macroscopic dynamical system: 

\begin{align*}
&C_m \frac{dr}{dt} = (b - \sum_Zs_{WZ})r_X + 2arv - (\pm \Delta)\frac{a}{C_m \pi}\\
&C_m \frac{dv}{dt} = -\frac{C_m^2 \pi^2}{a}r^2 + av^2 + c  - u + bv_X + I_{ext} + \overline{\eta}
\end{align*}

Since the firing rate always has to be non-negative, we needed to evaluate the closed integral contour containing the pole of $g(\eta)$ in the lower half of $\eta$ plane, i.e., $\overline{\eta} - i \Delta$. Until now, we considered the integral contour in both the upper and lower half of the $\eta$. This is because the Lorentzian variables $x$ and $y$ have no physical meaning. Therefore we could not make any conclusions regarding which contour to consider when using the residue theorem to solve \eqref{eq:r2} and \eqref{eq:v2} until now. 

We have that a mean-field reduction of a population of interacting conductance-based Izhikevich two-dimensional QIF neurons is given by :

\label{all:mean_field}
\begin{align}
 C_m \frac{dr_W}{dt} &= (b - \sum_Z s_{WZ})r_W + 2ar_Wv_W\nonumber \\
                    &+ \Delta\frac{a}{C_m \pi}\\
C_m \frac{dv_W}{dt} &= -\frac{C_m^2 \pi^2}{a}r_W^2 + av_W^2 + c  - u^W \nonumber \\
  &+ (b - \sum_Z s_{WZ})v_W + I_{ext} \nonumber \\
  &+ \sum_Z E_r^Z + \overline{\eta} 
\end{align}

with 

\begin{equation}
\frac{ds_{WZ}}{dt} = - \frac{s_{WZ}}{\tau_s} + p_{WZ}r_Z
\end{equation}

and where $u^W$ is the mean recovery current given by

\begin{align}
 \frac{du^W}{dt} & = \int_{\eta} \int_{v} \int_u u\frac{\partial}{\partial t} \rho(V|\eta)g(\eta) du dv d\eta \nonumber \\
                 &=  \alpha (\beta(v^W - V_r) - u^W) + u_{jump}r_W
\end{align}

\section{Validity of the Lorentzian ansatz}
\label{ap:LA}

Previous work by \cite{Montbrio} shows  how the dynamics of a class of QIF neurons generally converges to the Ott-Antonsen ansatz (OA) manifold.  This is known has the Lorentzian ansatz (LA). In this section, we clarify why the Lorentzian ansatz holds for the ensembles of QIF neurons here considered.

We start by introducing the following transformation:

\begin{equation}
V_i^W =tan\frac{\theta_i^W}{2}
\end{equation}

Then, Equation \ref{eq:V_QIF} transforms into:

\begin{align}
C_m \frac{d\theta_i^W}{dt} & = a(1 - cos \theta_i^W) + (c - u^W + \eta_i + \sum_Z s_{WZ}E_r^Z \nonumber \\
& + I_{ext}) + (b - \sum_Z s_{WZ})sin \theta_i^W
\end{align}

Note that $V=\pm \infty$ corresponds to $\theta = \pm \pi$.

According to the Ott-Antonsen ansatz \citep{ott2008low}, in the thermodynamic limit, the dynamics of a class of systems

\begin{equation}
\label{eq:Class_OA}
\frac{d \theta}{dt} = \Omega(\eta,t) + Im(H(\eta,t)e^{-i\theta})
\end{equation}

converges to the OA manifold

\begin{equation}
\tilde{\rho}(\theta | \eta, t) = \frac{1}{2\pi} Re[\frac{1 + \alpha(\eta,t)e^{i\theta}}{1 - \alpha(\eta,t)e^{i\theta}}]
\end{equation}

where the function $\alpha(\eta,t)$ is related to $w(\eta,t)=x(\eta,t) + iy(\eta,t)$ as

\begin{equation}
\alpha(\eta,t) = \frac{1 - w(\eta,t)}{1 + w(\eta,t)}
\end{equation}

Noticing that in the new variable $\theta^W$ our system belongs to the class \ref{eq:Class_OA} with $\Omega(\eta,t) = a + c + \sum_Zs_{WZ}E_r^Z + I_{ext} + \eta - u^W$ and $H(\eta,t)=(-b+\sum_Z s_{WZ}) + i(a - c - \sum_Zs_{WZ}E_r^Z - I_{ext} - \eta + u^W)$, we infer that it converges to:

\begin{widetext}
\begin{equation}
 \tilde{\rho}(\theta | \eta, t) = \frac{1}{\pi} Re[\frac{1+ xtan^2(\frac{\theta}{2}) + ytan(\frac{\theta}{2}) + i(ytan^2(\frac{\theta}{2}) + (1-x)tan(\frac{\theta}{2}) )}{tan^2(\frac{\theta}{2}) + x - ytan(\frac{\theta}{2}) + i( y - (1-x)tan(\frac{\theta}{2}) )}]
\end{equation}
\end{widetext}

Therefore, in the original variable $V^X$, our system converges to:

\begin{widetext}
\begin{equation}
 \rho(V^W| \eta, t) = \frac{1}{\pi} Re[\frac{1+ x(V^W)^2 + yV^W + i(y(V^W)^2 + (1-x)V^W )}{(V^W)^2 + x - yV^W + i( y - (1-x)V^W )}]
\end{equation}
\end{widetext}

After some algebraic manipulations, we recover the LA \eqref{rho}

\begin{align}
\rho(V^W|\eta, t) = \frac{1}{\pi} \frac{x(\eta,t)}{(V^W - y(\eta,t))^2 + x(\eta,t)^2}
\end{align}

The LA ansatz solves the continuity equation exactly, making the system amenable to theoretical analysis. In section \ref{subsec:sim}, we show that these solutions agree with the numerical simulations of the original QIF neurons, further validating the application of the LA.

\section{Mean-field reduction of theta neuron population}
\label{sec:burst_MF}

To get the mean-field reduction of a population of theta neurons, we will make use of the relation $W=\pi r + iv$ and $W = \frac{1-Z^*}{1 + Z^*}$ to map the macroscopic quantities of the population of QIF neurons (r, v, u) to the Kuramoto order parameter Z. Note that the conformal map  $W = \frac{1-Z^*}{1 + Z^*}$ is valid for the mapping $V=tan(\frac{\theta}{2})$ between QIF and theta neurons.

To ensure that, in the end, we will have a population of bursting neurons, we start by finding a bursting regime of theta neurons, such as:

\begin{subequations}
\label{eq:theta_neuron_burst}
\begin{align}
& \frac{d\theta_i}{dt} = 2( (1 - cos(\frac{\theta_i}{2})) + (1+cos(\frac{\theta_i}{2}) )(I - u_i ) )\\
&\frac{du_i}{dt} = \alpha ( \beta(  1 + \frac{tan(\frac{\theta_i}{4})}{1 + 2( 1 + tan(\frac{\theta_i}{4})^2  )}  ) - u_i )
\end{align}
\end{subequations}

The mapping $\theta_i = 2 tan^{-1}(V_i)$ will result in a QIF model ($V_i, u_i$) that it cannot be solved analytically in $V$, which means we cannot get the description of the macroscopic variable (r, v, u). The transformation $\theta_i = 2 tan^{-1}(V_i)$ will result in a integrable bursting version of ($V_i, u_i$), but then the conformal map $W = \frac{1-Z^*}{1 + Z^*}$ is no longer valid. To overcome this problem, we consider the Kuramoto order parameter $Z$ in the form $Z = |Z|exp(-i \Psi)$ and $V = tan(\frac{\Psi}{2})$, apply the mean-field reduction method above described, and use the change of variables $\Psi = 2 \theta$ to obtain a macroscopic description of a population of bursting theta neurons.

We start by obtaining the macroscopic description of the modified Izhikevich model

\begin{subequations}
\label{eq:new_QIF}
\begin{align}
&\frac{dV_i}{dt} = V_i^2 - u_i + I_i \\
&\frac{du_i}{dt} = \alpha ( \beta(1+ \frac{V_i}{2(V_i^2+1)+1} ) -u_i )
\end{align}
\end{subequations}

As a reminder, if we apply a change of variables $V_i = tan(\theta_i)$ to equations \eqref{eq:new_QIF} we get a model for bursting theta neurons.

Following the same procedure as in Appendix A, we get

\begin{subequations}
\label{eqs:MF_Vu_modified}
\begin{align}
\frac{dr}{dt} =& 2rv + bR + \frac{\Delta}{\pi}  \\
\frac{dv}{dt} =& -\pi^2r^2 + v^2  + c - u + bv + \bar{\eta} +I \\
\frac{du}{dt} =& \alpha (\beta (1 +  \frac{-2v \pi (3+2v^2 + 2r^2 \pi^2) + 4\sqrt{6}vr \pi^2}{2\pi ( 4v^2 + (3-2r^2\pi^2)^2 + 4v^2(3+2r^2\pi^2) )}  ) \nonumber \\
&- u)
\end{align}
\end{subequations}

where the mean recovery current $u$ was estimated according to the following equation: 

\begin{align*}
\frac{du}{dt}  =& \int_{\eta} \int_{v} \int_u u\frac{\partial}{\partial t} \rho(V|\eta)g(\eta) du dv d\eta \\
               =& \alpha [ \beta (1 + \int_{\eta} \int_v \frac{V}{2(V^2 + 1) + 1} \rho(V|\eta) g(\eta) dv d\eta) - u ] \\
               =& \alpha [ \beta (1 + \int_{\eta} \int_v \frac{x}{\pi}\frac{V}{(2(V^2 + 1) + 1) ((V-y)^2 + x^2 } g(\eta) dv d\eta)\nonumber \\
              &- u ]
\end{align*}

Then, we use the relations $W = \pi r + i v$ and $W = \frac{1 - Z^*}{1 + Z^*}$ to obtain the dynamics of the macroscopic variable $Z = |Z|exp(-i\Psi)$ and $u(Z,t)$. Using the former, we can re-write equations \eqref{eqs:MF_Vu_modified} in terms of $W$:\\

\begin{widetext}
\begin{subequations}
\begin{align}
&\frac{dW}{dt} = bW + \Delta + i(I + \bar{\eta} + c - u ) - iW^2 \\
&\frac{du}{dt} =  \alpha (\beta (1 +  \frac{-2 Im(W) \pi (3+2Im(W)^2 + 2Re(W)^2) + 4\sqrt{6} Re(W) Im(W) \pi}{2\pi ( 4Im(W)^2 + (3-2Re(W)^2)^2 + 4Im(W)^2(3+2Re(W)^2) )}  )  - u) 
\end{align}
\end{subequations}
\end{widetext}

Using the conformal mapping $W = \frac{1 - Z^*}{1+Z^*}$, we obtain:

\begin{widetext}
\begin{subequations}
\begin{align}
&\frac{dZ^*}{dt} = -\frac{1}{2}( b(1-Z^*) + \Delta(1+Z^*)^2 + i(1+Z^*)^2(I + \bar{\eta} + c - u ) - i(1-Z^*)^2 )\\
&\frac{du}{dt} =  \alpha (\beta (1 +  \frac{-2 Im(W) \pi (3+2Im(W)^2 + 2Re(W)^2) + 4\sqrt{6} Re(W) Im(W) \pi}{2\pi ( 4Im(W)^2 + (3-2Re(W)^2)^2 + 4Im(W)^2(3+2Re(W)^2) )}  )  - u) 
\end{align}
\end{subequations}
\end{widetext}

with

\begin{align*}
& Re(W) = \frac{1 - Re(Z^*)^2 - Im(Z^*)^2}{(1 + Re(Z^*))^2 + Im(Z^*)^2} \\
& Im(W) = -\frac{2 Im(Z^*)}{(1 + Re(Z^*))^2 + Im(Z*)^2} 
\end{align*}

Considering $Z$ in the polar form $Z = |Z|exp(-i\Psi)$ we get that:

\begin{widetext}
 \begin{subequations}
\label{eq:burst_MF_app}
\begin{align}
\frac{d|Z|}{dt} & =  \frac{1}{2}(   sin(\Psi)(\bar{\eta} + I_{ext} - u - 1) - \Delta cos(\Psi) (1 + |Z|^2) - 2|Z|\Delta + |Z|^2 sin(\Psi)(1 - \bar{\eta} - I_{ext} + u)   )  \\
\frac{d\Psi}{dt} & =  \frac{1}{2} (\frac{cos(\Psi)}{2}(\bar{\eta} + I_{ext} - u - 1) + \frac{\Delta}{2}sin(\Psi)(1 - |Z|^2) + 2(\bar{\eta} + I_{ext} - u) + 2 + |Z|cos(\Psi)(\bar{\eta} + I_{ext} - u - 1) ) \\
\frac{du}{dt}  & =\alpha (\beta (1 +  \frac{-2 Im(W) \pi (3+2Im(W)^2 + 2Re(W)^2) + 4\sqrt{6} Re(W) Im(W) \pi}{2\pi ( 4Im(W)^2 + (3-2Re(W)^2)^2 + 4Im(W)^2(3+2Re(W)^2) )}  )  - u)  
\end{align}
\end{subequations}
\end{widetext}

with

\begin{align*}
& Re(W) = \frac{1 - (|Z|cos(\Psi))^2 - (|Z|sin(\Psi))^2}{(1 + |Z|cos(\Psi))^2 + (|Z|sin(\Psi))^2}  \\
& Im(W) = -\frac{2 |Z|sin(\Psi)}{(1 + (|Z|sin(\Psi))^2 + (|Z|sin(\Psi))^2} 
\end{align*}

\begin{subequations}
\label{eq:burst_MF_final}
\begin{align}
\frac{d|Z|}{dt} & =  \frac{1}{2}(   sin(\frac{\theta}{2})(\bar{\eta} + I_{ext} - u - 1) - \Delta cos(\frac{\theta}{2}) (1 + |Z|^2) \nonumber \\
                             & - 2|Z|\Delta + |Z|^2 sin(\frac{\theta}{2})(1 - \bar{\eta} - I_{ext} + u)   )  \\
\frac{d\theta}{dt} & =   \frac{cos(\frac{\theta}{2})}{2}(\bar{\eta} + I_{ext} - u - 1) + \frac{\Delta}{2}sin(\frac{\theta}{2})(1 - |Z|^2) \nonumber \\
                            & + 2(\bar{\eta} + I_{ext} - u) + 2 \nonumber \\ & + |Z|cos(\frac{\theta}{2})(\bar{\eta} + I_{ext} - u - 1) \\
\frac{du}{dt}  & = \alpha( \beta(1 - \frac{B( 3-2\sqrt{6}A + 2A^2 + 2B^2 )}{8B^2 (2 + B^2) + (3 - 2A^2)^2}) -u^Y) 
\end{align}
\end{subequations}

with $A = \frac{1 - |Z|^2}{1 + |Z|^2 + 2|Z| cos(\frac{\theta}{2})}$ and $B = \frac{ -2|Z|sin(\frac{\theta}{2})}{ 1 + |Z|^2 + 2|Z| cos(\frac{\theta}{2}) }$.

One can use the variable transformation $V=tan(\frac{\theta}{4})$ from the start and avoid this last step. However, in that case, we would have to find the new conformal mapping that relates $W$ and $Z$. 

Please note that 1) there are other descriptions of $(\theta, u)$ that, with the right parameters, show bursting behavior, and 2) one could obtain the mean-field description of the two-variable theta model by integrating $\frac{d\theta}{dt}$ and $\frac{du}{dt}$ in $\theta$ and $\eta$ directly. We have chosen the option that seemed the easiest to implement (despite not being the most straightforward).


\bibliography{apssamp}

\end{document}